\documentstyle[12pt]{article}
\catcode`\@=11

\@addtoreset{equation}{section}
\def\theequation{\arabic{section}.\arabic{equation}}

\catcode`\@=11

\def\section{\@startsection{section}{1}{\z@}{3.5ex plus 1ex minus
   .2ex}{2.3ex plus .2ex}{\large\bf}}

\newskip\humongous \humongous=0pt plus 1000pt minus 1000pt

\newif\ifdtup

\def\eqnarray{\stepcounter{equation}\let\@currentlabel=\theequation
    \global\@eqnswtrue
    \global\@eqcnt\z@\tabskip\@centering\let\\=\@eqncr
    $$\halign to \displaywidth\bgroup\@eqnsel\hskip\@centering
      $\displaystyle\tabskip\z@{##}$&\global\@eqcnt\@ne
       \hfil${{}##{}}$\hfil
      &\global\@eqcnt\tw@ $\displaystyle\tabskip\z@{##}$\hfil
       \tabskip\@centering&\llap{##}\tabskip\z@\cr}
\def\lefteqn#1{\hbox to 4\arraycolsep{$\displaystyle #1$\hss}}
\def\thesection{\arabic{section}.}

\def\appendix{\setcounter{section}{0}
        \def\thesection{Appendix.}
        \def\theequation{\Alph{section}.\arabic{equation}}}

\long\def\@makefntext#1{\parindent 0cm\noindent
\hbox to 1em{\hss$^{\@thefnmark}$}#1}
\def\IR{{\hbox{{\rm I}\kern-.2em\hbox{\rm R}}}}
\def\IH{{\hbox{{\rm I}\kern-.2em\hbox{\rm H}}}}
\def\IC{{\ \hbox{{\rm I}\kern-.6em\hbox{\bf C}}}}
\def\IZ{{\hbox{{\rm Z}\kern-.4em\hbox{\rm Z}}}}
\def\rref#1{(\ref{#1})}
\newcommand{\beq}{\begin{equation}}
\newcommand{\eeq}{\end{equation}}
\newcommand{\Vol}{\hbox{\em Vol}\,}
\newcommand{\A}{{\bar A}}
\newcommand{\g}{{\bar g}}
\newcommand{\D}{{D}}

\newcommand{\q}{{\bar q}}
\newcommand{\Res}{{\bf Z}_p^{*}}
\begin{document}
%
%
%
%
\def\citen#1{%
\edef\@tempa{\@ignspaftercomma,#1, \@end, }
\edef\@tempa{\expandafter\@ignendcommas\@tempa\@end}%
\if@filesw \immediate \write \@auxout {\string \citation {\@tempa}}\fi
\@tempcntb\m@ne \let\@h@ld\relax \let\@citea\@empty
\@for \@citeb:=\@tempa\do {\@cmpresscites}%
\@h@ld}
%
\def\@ignspaftercomma#1, {\ifx\@end#1\@empty\else
   #1,\expandafter\@ignspaftercomma\fi}
\def\@ignendcommas,#1,\@end{#1}
%
%
\def\@cmpresscites{%
 \expandafter\let \expandafter\@B@citeB \csname b@\@citeb \endcsname
 \ifx\@B@citeB\relax 
    \@h@ld\@citea\@tempcntb\m@ne{\bf ?}%
    \@warning {Citation `\@citeb ' on page \thepage \space undefined}%
 \else
    \@tempcnta\@tempcntb \advance\@tempcnta\@ne
    \setbox\z@\hbox\bgroup 
    \ifnum\z@<0\@B@citeB \relax
       \egroup \@tempcntb\@B@citeB \relax
       \else \egroup \@tempcntb\m@ne \fi
    \ifnum\@tempcnta=\@tempcntb 
       \ifx\@h@ld\relax 
          \edef \@h@ld{\@citea\@B@citeB}%
       \else 
          \edef\@h@ld{\hbox{--}\penalty\@highpenalty \@B@citeB}%
       \fi
    \else   
       \@h@ld \@citea \@B@citeB \let\@h@ld\relax
 \fi\fi%
 \let\@citea\@citepunct
}
%
\def\@citepunct{,\penalty\@highpenalty\hskip.13em plus.1em minus.1em}%
%
%
\def\@citex[#1]#2{\@cite{\citen{#2}}{#1}}%
%
%
\def\@cite#1#2{\leavevmode\unskip
  \ifnum\lastpenalty=\z@ \penalty\@highpenalty \fi 
  \ [{\multiply\@highpenalty 3 #1
      \if@tempswa,\penalty\@highpenalty\ #2\fi 
    }]\spacefactor\@m}
\let\nocitecount\relax  
%
\begin{titlepage}
\vspace{.5in}
\begin{flushright}
UCD-92-16\\                             
hep-th/9206103                          
June 1992\\
\end{flushright}
\vspace{.5in}
\begin{center}
{\Large\bf
The Sum over Topologies\\
in Three-Dimensional Euclidean Quantum Gravity}\\
\vspace{.4in}
{S.~C{\sc arlip}\footnote{\it email: carlip@dirac.ucdavis.edu}\\
       {\small\it Department of Physics}\\
       {\small\it University of California}\\
       {\small\it Davis, CA 95616}\\{\small\it USA}}
\end{center}

\vspace{.5in}
\begin{center}
{\large\bf Abstract}
\end{center}
{\small In Hawking's  Euclidean path integral approach to quantum gravity,
the partition function is computed by summing contributions from all
possible topologies.  The behavior such a sum can be estimated in three
spacetime dimensions in the limit of small cosmological constant.  The
sum over topologies diverges for either sign of $\Lambda$, but for
dramatically different reasons: for $\Lambda>0$, the divergent behavior
comes from the contributions of very low volume, topologically complex
manifolds, while for $\Lambda<0$ it is a consequence of the existence of
infinite sequences of relatively high volume manifolds with converging
geometries.  Possible implications for four-dimensional quantum gravity
are discussed.}
\end{titlepage}
\addtocounter{footnote}{-1}

In Hawking's Euclidean path integral approach to quantum gravity
\cite{Hawking}, the partition function is computed as a weighted sum
over topologies, with the contribution of each four-manifold $M$
determined by a path integral over Riemannian (positive definite) metrics.
This program has not been easy to implement --- general relativity
is nonrenormalizable, the action is apparently unbounded below, and the
relevant topologies cannot be classified --- but it seems as plausible
an approach as any for quantizing general relativity.  It would therefore
be useful to have a simple model in which more explicit calculations could
be made.

Three-dimensional gravity provides such a model.  Three-dimensional
solutions of the empty space Einstein equations necessarily have constant
Riemann curvature, so the classical theory is dramatically simplified.  At
the quantum level, the three-dimensional model is renormalizable, and its
relationship to Chern-Simons theories makes a systematic perturbation
expansion possible.  The three-dimensional path integral may also have a
more direct (3+1)-dimensional interpretation in terms of the finite
temperature partition function.

The goal of this paper is to evaluate the sum over three-dimensional
topologies to first order in the loop expansion.  We shall see that
the behavior of this sum depends strongly on the sign of the
cosmological constant.  For $\Lambda>0$, the sum is dominated by very
low volume manifolds with complicated topologies, giving a highly
nonclassical behavior.  For $\Lambda<0$, on the other hand, the main
contributions come from infinite sequences of relatively high volume
manifolds whose geometries converge to those of incomplete (cusped)
manifolds.  The sensitivity to the sign of $\Lambda$ comes as something
of a surprise, and suggests that even a very small cosmological constant
can have dramatic quantum mechanical effects.

\section{The Partition Function}
\setcounter{footnote}{0}

For a compact three-manifold $M$, the ``Wick rotated'' gravitational path
integral takes the form
\beq
Z_M = \int [dg] \exp\left\{ {1\over16\pi G}\int_M d^3\!x
    \sqrt{g}(R[g]-2\Lambda)\right\}
    \ ,
\label{1x1}
\eeq
where the integration is over all Riemannian metrics $g$ on $M$, $R[g]$ is
the scalar curvature, and $\Lambda$ is the cosmological constant.  The
extrema of the action are Einstein spaces, that is, manifolds and metrics
for which
\beq
R_{ik}[\g] = 2\Lambda \g_{ik} \ .
\label{1x2}
\eeq
In three dimensions, the Ricci tensor completely determines the full
curvature; in particular, an Einstein space always has constant curvature.
This means that an extremum $(M,\g)$ of the action must be an elliptic
($\Lambda>0$), hyperbolic ($\Lambda<0$), or flat ($\Lambda=0$) manifold.
For now, we shall take $\Lambda\ne 0$; the flat case will be discussed
briefly in the conclusion.

In the neighborhood of an extremum, standard techniques may be used to
evaluate the path integral \rref{1x1} in the saddle point (or
``semiclassical'') approximation.  One obtains an expression of the form
\beq
Z_M =  \left| \pi_0(\mbox{\it Diff}\,M )\right|^{-1}
      \sum_{\mbox{\scriptsize extrema}}\D_M \exp\left\{
      \mbox{sign}(\Lambda) {\Vol(M)\over\,4\pi G|\Lambda|^{1/2}}\right\} \ ,
\label{1x3}
\eeq
where $\Vol(M)$ is the volume of $M$ with the constant curvature metric
rescaled to have curvature $\pm 1$.  $|\pi_0(\mbox{\it Diff}\,M)|$
is the order of the mapping class group of $M$, that is, the number of
equivalence classes of diffeomorphisms of $M$ that cannot be smoothly
deformed to the identity.  This factor is needed to compensate for the
overcounting of physically equivalent metrics that differ by such ``large''
diffeomorphisms; in the saddle point approximation, it can be omitted as
long as we include only one extremum from each diffeomorphism class of
metrics.  The prefactor $\D_M$ is a combination of determinants coming
from small fluctuations around $\g$ and from gauge-fixing, and can be
computed explicitly by taking advantage of the connection between
three-dimensional gravity and Chern-Simons theory \cite{Witten}.  The
result may be summarized as follows.

Any compact, constant curvature three-manifold $M$ can be expressed as a
quotient space
\beq
M = {\widetilde M}/\Gamma \ ,
\label{1x4}
\eeq
where the universal covering space $\widetilde M$ is either the three-sphere
($\Lambda>0$) or hyperbolic three-space ($\Lambda<0$), and $\Gamma$ is a
discrete group of isometries of $\widetilde M$.  Denote the full group of
isometries of $\widetilde M$ by $G$; $G$ will be either SO(4) ($\Lambda>0$)
or SL($2,\IC$) ($\Lambda<0$).  The group $\Gamma$ acts on $\widetilde M$ by
isometries and on $G$ by the adjoint action, and one can construct a
$G$-bundle
\beq
E_M = (G\times \widetilde M)/\Gamma \ ,
\label{1x5}
\eeq
where the quotient is by this combined action.  This bundle arises in the
theory of geometric structures \cite{Goldman}, and admits a natural flat
connection $\A$ with a holonomy group isomorphic to $\Gamma$.  The
covariant derivative $D_\A$ then determines a Ray-Singer torsion
\beq
T(E_M) = {(\mbox{det}'\Delta_0)^{3/2}\over(\mbox{det}'\Delta_1)^{1/2}} \ ,
\label{1x5a}
\eeq
where $\Delta_k$ is the Laplacian formed from $D_A$ acting on
$(\mbox{Ad}\,G)$-valued $k$-forms. The definition of such a Laplacian
requires a metric; we choose the extremal metric on $M$ scaled to curvature
$\pm 1$.\footnote{The Ray-Singer torsion is independent of this metric only
when none of the Laplacians appearing in its definition have zero-modes.}
Denote the centralizer of $\Gamma$ in $G$ by $Z(\Gamma)$, and let
$\Vol(Z(\Gamma))$ be its volume with respect to the Haar measure.  Up to
an overall constant independent of the topology of $M$, the prefactor is
then
\beq
\D_M =
\left(G|\Lambda|^{1/2}|\Gamma|\right)^{\mbox{\scriptsize dim}\,Z(\Gamma)/2}
      \Vol(Z(\Gamma))^{-1} T(E_M)^{1/2} \ ,
\label{1x6}
\eeq
where $|\Gamma|$ denotes the order of the group $\Gamma$.

The torsion dependence of \rref{1x6} follows from two results of Witten.
In reference \cite{Witten}, it is shown that three-dimensional gravity
can be rewritten as a Chern-Simons theory for the gauge group $G$.  An
extremum of the action thus determines a flat connection, with a
corresponding bundle given by \rref{1x5}.  The saddle point approximation
can then be extracted from the results obtained in \cite{Witten_Jones}
for arbitrary Chern-Simons theories (see also \cite{WBN} for a careful
extension to noncompact gauge groups, and \cite{Carlip_entropy} for
an application to gravity).  Note that for Euclidean quantum gravity,
there is no imaginary contribution to the effective action, i.e.,
no complex phase in \rref{1x6}.  This is most easily understood by
noting that the phase in a Chern-Simons path integral \cite{Witten_Jones}
comes from regulating integrals of the form
$$\int {dx\over\sqrt{2\pi}} e^{-i\lambda x^2/2} \ ,$$
while for Euclidean gravity the exponent in \rref{1x1} is real and the
corresponding integrals are absolutely convergent.

The remaining terms in \rref{1x6} come from ghost zero-modes, that is,
from the subgroup ${\cal G}_0$ of gauge transformations that leave the
extremal connection $\A$ fixed.  These factors are of the form
$\Vol({\cal G}_0)^{-1}$, appropriately normalized, and have their
origin in the standard Faddeev-Popov gauge-fixing procedure: the gauge
orbit through $\A$ is parameterized by ${\cal G}/{\cal G}_0$ rather than
the full gauge group $\cal G$, so when the volume $\Vol({\cal G})$ is
factored out of the partition function, a term $\Vol({\cal G}_0)$
remains in the denominator.  In our case, ${\cal G}_0$ is isomorphic
to the centralizer $Z(\Gamma)$.  Indeed, gauge transformations act
on the holonomy by conjugation, so only transformations that commute
with every $\gamma\in\Gamma$ can leave $\A$ invariant, while conversely,
any $g\in Z(\Gamma)$ can be extended to all of $M$ by parallel transport
to define a gauge transformation that fixes $\A$.

To obtain the proper normalization of the volume $\Vol({\cal G}_0)$,
recall that the path integral measure is fixed by the condition
\beq
\int [d\epsilon]
     \exp\left\{ -k\int_M \mbox{Tr}\,\epsilon\wedge *\epsilon \right\}
     = 1 \ .
\label{1x8}
\eeq
Here $k=1/(64\pi G|\Lambda|^{1/2})$ is the constant multiplying the
Chern-Simons action, and the Hodge dual $*$ is defined in terms of the
metric of constant curvature $\pm1$ on $M$ to ensure consistency with
\rref{1x5a}. Let us choose as a basis an orthonormal set of
$(\mbox{Ad}\,G)$-valued eigenfunctions of the Laplacian on $\widetilde M$,
\beq
\Delta_0 u_\alpha = \lambda_\alpha u_\alpha \ ,\qquad
\int_{\widetilde M} \mbox{Tr}\,\left(u_\alpha\wedge*u_\beta \right)
    = \delta_{\alpha\beta} \ .
\label{1x9}
\eeq
A corresponding basis for sections of $E_M$ is given by
\beq
\left\{ u_\alpha\ |\ u_\alpha(\gamma x) = \gamma^{-1}u_\alpha(x)\gamma
   \quad \forall \gamma\in\Gamma \right\} \ .
\label{1x9a}
\eeq
If we now write $\epsilon = \epsilon^\alpha u_\alpha$, it is evident from
\rref{1x8} that the measure must take the form
\beq
[d\epsilon] =
   \prod_\alpha\, \left({k\over2\pi|\Gamma|}\right)^{1/2}
   d\epsilon^\alpha \ .
\label{1x10}
\eeq
The factor $|\Gamma|$ comes from the fact that we are integrating over
$M = \widetilde M/\Gamma$ rather than $\widetilde M$.\footnote{Up to an
overall constant, $|\Gamma|$ can be replaced by $\Vol(M)^{-1}$.  This
permits a simple generalization to the case in which $\Gamma$ has infinite
order.}  The generators of ${\cal G}_0\approx Z(\Gamma)$ are now the
zero-modes $D_\A u_\alpha = 0$.  Integrating the measure \rref{1x10} over
this space of zero-modes and using the definition of $k$, we obtain
\beq
\Vol({\cal G}_0) = \mbox{const.}\,
\left(G|\Lambda|^{1/2}|\Gamma|\right)^{-\mbox{\scriptsize dim}\,Z(\Gamma)/2}
\Vol(Z(\Gamma)) \ ,
\label{1x11a}
\eeq
giving the corresponding factor in \rref{1x6}.

In principle, the Chern-Simons formulation also makes it possible to
calculate higher order corrections to the saddle point approximation.
Such calculations have not yet been carried out (although exact results
for SU(2) Chern-Simons theory might be extendible to SO(4)), but we know
from \cite{Witten_Jones} that these corrections will be suppressed by
factors of $1/k\sim G|\Lambda|^{1/2}$.  For small cosmological constant,
the semiclassical approximation \rref{1x6} should therefore be reliable.

\section{Positive Cosmological Constant}
\setcounter{footnote}{0}

Let us now consider the case $\Lambda>0$ in more detail.  We are faced with
several tasks: we must classify the possible extremal manifolds and metrics
$(M,\g)$, or equivalently the discrete groups of isometries $\Gamma$; for
each such manifold, we must evaluate the volume $\Vol(M)$ and the Ray-Singer
torsion $T(E_M)$; and we must compute the remaining normalization factors
in \rref{1x6}.  Fortunately, all of these tasks are manageable.

Elliptic three-manifolds --- three-manifolds of constant positive curvature
--- were classified by Seifert and Threlfall in 1930 \cite{SeifThrel}.  A
recent summary is given by Wolf \cite{Wolf} (see also \cite{Witt,Thomas}).
Reference \cite{Wolf} contains an explicit description of each admissible
group $\Gamma$ of isometries of $S^3$.  Given such a group, the volume
$\Vol(M)$ is simply
\beq
\Vol(M) = {2\pi^2/|\Gamma|} \ .
\label{2x1}
\eeq
($2\pi^2$ is the volume of the three-sphere of constant curvature $+1$.)
Note that for topologically complicated three-manifolds, the order
$|\Gamma|$ becomes large, $\Vol(M)$ is small, and the exponent in \rref{1x3}
approaches zero.  We shall see later that the behavior of manifolds with
$\Lambda<0$ is dramatically different.

The Ray-Singer torsion for $M$ can be calculated from the results of Ray
\cite{Ray}:\footnote{Strictly speaking, Ray's paper only deals with the
torsion for one-dimensional representations of $\Gamma$, but our case ---
in which the adjoint representation is needed --- is an easy generalization;
see, for instance, \cite{Lott}.  Note that Ray's normalization differs
from that of \rref{1x5a}.}
\beq
T(E_M) =
\prod_{k=1}^{|\Gamma|} \left| 1-e^{2\pi i k/|\Gamma|} \right|^{{1\over2}A_k}
\ ,
\label{2x2}
\eeq
where
\beq
A_k = {1\over|\Gamma|} \sum_{g\in\Gamma} \mbox{Tr}\,\rho_A(g)
      \mbox{Tr}\,\rho_0(g^k) \ .
\label{2x3}
\eeq
In this last expression, $\rho_A(g)$ is the adjoint representation of $g$,
while $\rho_0(g)$ is the representation of $g$ as a group of rotations in
$\IR^4$.  In particular, if $g$ is a rotation of period $p$, one can pick
complex coordinates $z_1$ and $z_2$ such that
\beq
g z_j = e^{2\pi i \nu_j/p} z_j \ ;
\label{2x4}
\eeq
then
\beq
\mbox{Tr}\,\rho_0(g^k) = 2\cos{2\pi k \nu_1\over p} +
   2\cos{2\pi k \nu_2\over p} \ .
\label{2x5}
\eeq
We also need the Haar volume $\Vol(Z(\Gamma))$; this is easily calculated
from Wolf's explicit representations of $\Gamma$ \cite{Wolf}.

To obtain some more explicit results, let us work out the path integral
for the case of cyclic fundamental groups $\Gamma\approx\IZ_p$.  The
corresponding three-manifolds are lens spaces $L_{p,q}$, where $q$
is an integer relatively prime to $p$ with $1\le q\le p$.  Two
lens spaces $L_{p,q}$ and $L_{p,\q}$ are topologically distinct unless
$q\q = 1\, (\mbox{mod}\, p)$.  In this aspect, lens spaces are exceptional
among elliptic three-manifolds, which are usually uniquely determined by
their fundamental groups.  One might therefore expect lens spaces to
dominate the partition function, since for a typical volume $\Vol(M) =
2\pi^2/p$, most of the saddle points contributing to the sum over topologies
will be lens spaces.  Moreover, as $p$ becomes large, the exponent in
\rref{1x3} will become unimportant, while the number of lens spaces will
grow.  One might thus anticipate that manifolds with large fundamental
groups dominate.  This will indeed prove to be the case.

In the representation \rref{2x4}, the holonomy group $\Gamma_{p,q}$ of
$L_{p,q}$ is generated by an element $g$ for which
\beq
\nu_1 = 1 \ ,\quad \nu_2 = q \ ,
\label{2x6}
\eeq
while of course $|\Gamma| = p$.  The evaluation of the traces in \rref{2x3}
is straightforward, and the sum over elements in $\Gamma$ reduces to a
sum over powers of $g$.  Using the identity
\beq
\sum_{j=0}^{p-1} e^{2\pi i m j/p}
  =\left\{ \begin{array}{ll} p&\mbox{if $m=0\pmod p$}\\
                             0&\mbox{otherwise}\end{array} \right. \ ,
\label{2x7}
\eeq
we find a Ray-Singer torsion (for $q\ne\pm 1\, (\mbox{mod}\, p)$)
\beq
T(E_{p,q})^{1/2} = 4\left(\cos{2\pi\over p} - \cos{2\pi q\over p}\right)
    \left(\cos{2\pi\over p} - \cos{2\pi \q\over p}\right) \ ,
\label{2x8}
\eeq
where $1\le \q\le p$ is the integer determined by $q\q = 1\,
(\mbox{mod}\, p)$.  Moreover, for $q\ne\pm 1\, (\mbox{mod}\, p)$ the
centralizer $Z(\Gamma_{p,q})$ consists of rotations that can be represented
as
\beq
g z_j = e^{2\pi i \theta_j} z_j \ ,
\label{2x9}
\eeq
so $\mbox{dim}\,Z(\Gamma_{p,q}) = 2$.  The Haar volume
$\Vol(Z(\Gamma_{p,q}))$ is a constant independent of $p$ and $q$, and
the factor $|\pi_0(\mbox{\it Diff}\,L_{p,q})|$ can be omitted from
\rref{1x3} as long as we count only one saddle point contribution per
lens space.

Combining these results, we find a partition function
\beq
Z_{L_{p,q}} = c\, G\Lambda^{1/2} p
    \left(\cos{2\pi\over p} - \cos{2\pi q\over p}\right)
    \left(\cos{2\pi\over p} - \cos{2\pi \q\over p}\right)
    \exp\left\{{\pi\over2G\Lambda^{1/2}p}\right\} \
\label{2x11}
\eeq
where $c$ is a constant independent of $p$ and $q$. Equation \rref{2x11}
will fail to hold when $q=\pm1\, (\mbox{mod}\, p)$; this case could
be dealt with separately, but it is unimportant in understanding the
qualitative behavior of the sum over topologies, so we shall simply
ignore it.

To study the asymptotic behavior of the sum over topologies, let us
first sum \rref{2x11} over integers $1\le q\le p$ relatively prime to $p$.
Such a set of integers is called a reduced system of residues for $p$,
and will be denoted, with a slight abuse of notation, by $\Res$.  This sum
will overcount topologies by approximately a factor of two, since
$L_{p,q}\approx L_{p,\q}$.  (It is easily checked that as $q$ ranges over
a reduced system of residues, so does $\q$, so $L_{p,q}$ will appear twice
unless $q=\q$, i.e., unless $q^2=1\, (\mbox{mod}\, p)$.)

The sum over $q$ can be written in terms of three standard functions in
number theory.  Euler's function $\phi(p)$ is the number of positive
integers $1\le q\le p$ relatively prime to $p$.  Ramanujan's sum is
\beq
c_p(m) = \sum_{q\in\Res} \exp\{2\pi i mq/p\} \ ,
\label{2x12}
\eeq
and Kloosterman's sum is
\beq
S(m,n,p) = \sum_{q\in\Res} \exp\{2\pi i(mq + n\q)/p\} \ ,
\label{2x13}
\eeq
with $\q$ defined as above.  It is now easy to check that
\begin{eqnarray}
\sum_{q\in\Res} Z_{L_{p,q}} = && c G\Lambda^{1/2} p \\
    &&\cdot\left(\phi(p)\cos^2{2\pi\over p} -2c_p(1)\cos{2\pi\over p}
         + 2(S(1,1,p) + S(1,-1,p))\right)
    \exp\left\{{\pi\over2G\Lambda^{1/2}p}\right\}\nonumber \ .
\label{2x14}
\end{eqnarray}
For large $p$, the asymptotic behavior of these quantities is known
\cite{Hardy,Estermann}:
\begin{eqnarray}
\phi(p)&\sim&{6p\over\pi^2} \nonumber\\
c_p(1)&\sim&1\\
S(1,\pm1,p)&\sim&p^{1/2+\epsilon} \ . \nonumber
\label{2x15}
\end{eqnarray}
Hence
\beq
\sum_{q\in\Res} Z_{L_{p,q}} \sim c' G\Lambda^{1/2} p^2
\label{2x16}
\eeq
for large $p$.  As anticipated, the sum over topologies is dominated
by manifolds with arbitrarily large fundamental groups, and correspondingly
small volumes.

There should be no serious difficulty in extending this analysis to
other elliptic manifolds.  Some preliminary observations support the
suggestion that the lens spaces give the most important contribution.  For
instance, if $\Gamma$ is a binary dihedral group $D^*_{4n}$, $S^3/\Gamma$
is a ``prism manifold.''  The centralizer $Z(\Gamma)$ is then
three-dimensional, so the $\Vol({\cal G}_0)$ term in the prefactor
\rref{1x6} goes as $|\Gamma|^{3/2}$, but there is only one prism manifold
for each value of $|\Gamma|$, so the overall growth is slower than
\rref{2x16}.  For $\Gamma = \IZ_m\times D^*_{2n+1}$, there is more than
one relevant representation of $\Gamma$ for each pair $(m,n)$, and may
therefore be more than one inequivalent manifold (I do not know which
representations give diffeomorphic manifolds).  In this case, however,
the centralizer $Z(\Gamma)$ is only one-dimensional, so the growth is
again likely to be slower than \rref{2x16}.

\section{Negative Cosmological Constant}
\setcounter{footnote}{0}

We next turn to the case $\Lambda<0$.  Classical extrema of the Einstein
action are now hyperbolic manifolds, manifolds of constant negative
curvature.  Unlike elliptic manifolds, hyperbolic manifolds have not
been completely classified, so a systematic computation is not
yet possible.  Nevertheless, some useful qualitative conclusions can be
drawn.

Let us first consider the exponential term in the partition function
\rref{1x3}.  For $\Lambda<0$, manifolds with high volumes are exponentially
suppressed.  In contrast to the elliptic case, more complicated fundamental
groups now generally give larger volumes: $\Gamma$ can be viewed
as a set of gluing instructions for tetrahedral cells of $M$, and more
complicated groups require more ``building blocks.''  One might therefore
expect the partition function to be dominated by low volume, topologically
simple manifolds.

Next, the $\Vol({\cal G}_0)$ factor in $D_M$ drops out of \rref{1x6}
when $\Lambda$ is negative.  Geometrically, the centralizer $Z(\Gamma)$
is a measure of the size of the isometry group of $M$, and generic
hyperbolic three-manifolds admit no continuous isometries.  Thus
$\mbox{dim}\,Z(\Gamma)$ usually vanishes, and $D_M$ depends only on
the Ray-Singer torsion.

But while large volumes tend to suppress the contributions of manifolds
with complicated topologies, there is a competing effect coming from
statistical weight, or ``entropy'' --- the {\em number} of manifolds
contributing to the partition function may peak near certain volumes,
possibly overcoming the exponential suppression.  This phenomenon is
discussed in \cite{Carlip_entropy} in a slightly different context; it
is demonstrated there that the path integral for manifolds with a single
boundary component is dominated by such peaks.

To compare the effects of ``action'' and ``entropy'' in \rref{1x3}, we must
first understand the distribution of volumes of hyperbolic three-manifolds.
The crucial result is due to Thurston \cite{Thurston}.  We start with a
definition: a cusp of a hyperbolic three-manifold $M$ is a neighborhood
of an embedded circle that is ``infinitely far away'' from the rest of the
manifold in the hyperbolic metric.  Topologically, a neighborhood of a cusp
is diffeomorphic to $T^2\times[t_0,\infty)$, where $T^2$ is a two-dimensional
torus. (The circle itself is not part of $M$, which is complete but not
compact.)  Metrically, we can take the upper half space model for $\IH^3$,
with the standard constant negative curvature metric
\beq
ds^2 = t^{-2}(dx^2 + dy^2 + dt^2) \ ;
\label{3x1}
\eeq
a neighborhood of a cusp then looks like a region
\beq
N = \left\{ (x,y,t): t>t_0, z\sim z+1, z\sim z+\tau\right\} \ .
\label{3x2}
\eeq
Although cusped manifolds are incomplete, their volumes are finite, since
the area of a toroidal cross section of $N$ shrinks exponentially as
the proper distance $(\log t)$ goes to infinity.

Thurston has shown that the set of volumes of compact hyperbolic
three-manifolds has accumulation points precisely at
the volumes of cusped manifolds.  Given any cusped
manifold $M_\omega$, one can construct an infinite family of hyperbolic
manifolds $M_{p,q}$, where $p$ and $q$ are relatively prime integers, by
means of a technique known as hyperbolic Dehn surgery.\footnote{Strictly
speaking, a finite number of relatively prime pairs $(p,q)$ must be
excluded.  Without loss of generality, we can also take $p$ to be positive.}
For $p^2+q^2$ large, there is a reasonable sense in which the geometry of
the manifolds $M_{p,q}$ can be said to converge to that of $M_\omega$. In
particular, the volumes $\Vol(M_{p,q})$ converge (from below) to
$\Vol(M_\omega)$.  Thurston shows that every accumulation point in the
set of volumes can be obtained in this manner.  Any important peak
in the partition function due to ``entropy'' is thus likely to
occur near a cusped manifold.

The behavior of the Ray-Singer torsion $T(E_M)$ under hyperbolic Dehn
surgery has been investigated in \cite{Carlip_entropy}.  The results may
be summarized as follows.  Given relatively prime integers $p$ and $q$,
define two new integers $r$ and $s$ by the condition $ps-qr=1$.  Then for
$p$ and $q$ sufficiently large, the torsion $T_{p,q}$ of $M_{p,q}$ is
\beq
T_{p,q}^{1/2} =  T_\omega^{1/2} \sin^2 \left({2\pi r\over p}\right)
+ O({1\over p}) \ ,
\label{3x3}
\eeq
where $T_\omega$ is a suitably defined torsion for the cusped manifold
$M_\omega$.  The behavior of the volumes $\Vol(M_{p,q})$ is also known
\cite{NZ}:
\beq
\Vol(M_{p,q}) = \Vol(M_\omega) + O({1\over p^2+q^2}) \ .
\label{3x3a}
\eeq
The partition function for the sum over topologies $M_{p,q}$ obtained by
hyperbolic Dehn surgery can thus be approximated as
\beq
Z \sim \sum_{(p,q)=1}  T_\omega^{1/2} \sin^2 {2\pi r\over p}
   \exp\left\{-{\Vol(M_\omega)\over4\pi G|\Lambda|^{1/2}}\right\} \ .
\label{3x4}
\eeq

As in the elliptic case, this sum can be partially evaluated in terms of
functions that occur in number theory.  Let us write $q = np + q'$, where
$1\le q'\le p$ is relatively prime to $p$, i.e., $q'\in\Res$.  It may be
shown that as $q'$ varies a reduced set of residues for $p$, so does
$r$.  The sum over $q$ in \rref{3x4} thus becomes
$$ \sum_n \sum_{r\in\Res} \sin^2 {2\pi r\over p} \ ,$$
which can be expressed in terms of $\phi(p)$ and $c_p(2)$.  One obtains an
asymptotic behavior of the form
\beq
Z \sim \sum_{n,p} cp T_\omega^{1/2}\exp\left\{-{\Vol(M_\omega)\over
   4\pi G|\Lambda|^{1/2}}\right\} \ ,
\label{3x5}
\eeq
which clearly diverges.

In contrast to the elliptic case, divergences in the sum over topologies
may now come from manifolds with fairly large volumes.  The smallest cusped
hyperbolic manifold has a volume $v_0\approx 1.01494$, and manifolds with
$m$ cusps have volumes
\beq
\Vol(M) \ge mv_0 \ ;
\label{3x6}
\eeq
the inequality is strict for $m>2$ \cite{Adams}.  It is not entirely clear
how to insert a cut-off in the sum \rref{3x5}, so it is difficult to
compare divergences coming from different cusped manifolds.  But
hyperbolic Dehn surgery can be performed independently on each cusp of an
$m$-cusped manifold, giving $m$ independent divergent sums, so it can be
plausibly argued that the main contributions will come from manifolds with
large numbers of cusps, and hence large volumes.

\section{Conclusions}
\setcounter{footnote}{0}

We have now seen that the sum over topologies in three-dimensional
Euclidean quantum gravity depends dramatically on the sign of the
cosmological constant.  For $\Lambda>0$, the behavior of the partition
function is highly nonclassical --- the main contributions come from
manifolds with very small volumes and very complex topologies.  For
$\Lambda<0$, the dominant contributions come from infinite sequences
of relatively high volume manifolds, although still with complicated
topologies.  In both cases, it is clear that the ``leading'' saddle
point is relatively unimportant.

The particular form of these partition functions certainly depends strongly
on the fact that we are working in three dimensions.  But the qualitative
conclusions may well generalize to four dimensions.  The saddle points
that contribute to the sum over topologies will again depend strongly
on the sign of $\Lambda$, and as in three dimensions, the exponential
dependence on the classical action will tend to suppress complicated
topologies much more strongly when $\Lambda<0$.  Moreover, as in three
dimensions, the statistical weight will be of key importance when
$\Lambda<0$, since the number of topologically distinct Einstein spaces
with negative curvature grows very rapidly with volume \cite{Gromov}.

Three important approximations have been made in these calculations.
First, only the one-loop contribution to the prefactor \rref{1x6} has been
computed.  For large cosmological constant, this is a significant omission.
But for $\Lambda$ small, higher loop corrections are suppressed by powers
of $G|\Lambda|^{1/2}$, and should be unimportant for any given topology.
Note, however, that the prefactor can involve powers of $|\Gamma|$,
so the two-loop contribution of a topologically complex manifold could be
comparable to the one-loop contribution of a simpler one.  This can only be
determined by a careful evaluation of the $|\Gamma|$ dependence of higher
loop corrections.

Second, we have not yet summed over all saddle points.  For $\Lambda$
positive, there should be no serious difficulty in extending these results
to arbitrary elliptic manifolds, and it is likely that the lens space
contributions will dominate.  For $\Lambda$ negative, the problem is more
difficult, since hyperbolic three-manifolds have not been classified.  It
would be particularly useful to find a way to compare the contributions of
the series \rref{3x5} for neighborhoods of different cusped manifolds.

Third, we have not taken into account the contributions of topologies for
which the Einstein equations have no classical solution.  In Yang-Mills
theory, the contributions of approximate ``multi-instanton'' saddle points
are very important.  The analog here would be the contribution of connected
sums $M = M_1\#\dots\#M_n$.  (The connected sum of $M_1$ and $M_2$ is
formed by cutting out a three-ball from each manifold and identifying the
resulting spherical boundaries.)  In pure Chern-Simons theory, the partition
function for a connected sum is \cite{Witten_Jones}
\beq
Z(M_1\#\dots\#M_n) = {Z(M_1)\cdot\dots\cdot Z(M_n)\over Z(S^3)^{n-1}} \ .
\label{4x1}
\eeq
Gravity is not quite a pure Chern-Simons theory, however --- because of the
mapping class group dependence of \rref{1x3}, we need an additional factor
of
\beq
{|\pi_0(\mbox{\it Diff}\,M_1)|\cdots|\pi_0(\mbox{\it Diff}\,M_n)|
\over |\pi_0(\mbox{\it Diff}\,(M_1\#\dots\#M_n))|} \ ,
\label{4x2}
\eeq
which counts the new diffeomorphisms that appear when one forms a connected
sum.  Such diffeomorphisms include permutations of identical summands
$M_i$ and $M_j$, giving a factor of $1/m!$ for each set of $m$ identical
manifolds in the connected sum.  If this were the only contribution, the
effect of ``multi-instantons'' would be simply to exponentiate the
``one-instanton'' partition function we have already investigated.  An
assumption similar to this is made in wormhole dynamics \cite{Coleman}.

In fact, however, connected summation introduces a large number of
new diffeomorphisms, and the factor \rref{4x2} is generally much smaller
than $1/m!$.  In addition to permutations of summands, a connected sum
admits ``slide diffeomorphisms,'' diffeomorphisms in which one factor is
dragged around a closed curve in another factor \cite{DeSa,Hendriks,Anez}.
As the order $|\Gamma_i|$ of the fundamental group of $M_i$ grows,
the number of such diffeomorphisms increases rapidly.  It is
therefore plausible that the ``multi-instanton'' contribution will fall off
quickly as the topology of each instanton becomes more complicated.  This
effect was considered for $M=S^2\times S^1$ in \cite{CarAlwis}, although
in the somewhat different framework of the Lorentzian path integral; a
similar phenomenon lies behind the ``sewing problem'' in string theory
\cite{Car_sew}.  Clearly, a more careful analysis would be desirable.

Finally, let us briefly address the case of vanishing cosmological
constant.  The saddle points will now be flat three-manifolds, and the
exponential in \rref{1x3} will vanish.  Flat three-manifolds are completely
classified \cite{Wolf} --- all are finitely covered by the torus --- and
there is nothing to prevent us from repeating the calculations above.  In
fact, we can do better: when $\Lambda=0$, the the partition function
$Z_M$ is given {\em exactly} in terms of an integral over Ray-Singer
torsions \cite{Witten_top}.  However, as Witten discusses in
\cite{Witten_top}, the resulting integral typically diverges, so it is
not easy to compare contributions from different topologies.

\vspace{2.5ex}
\begin{flushleft}
\large\bf Acknowledgements
\end{flushleft}

I would like to thank Colin Adams, Scott Axelrod, and Walter Carlip for
useful advice.  This research was supported in part by the U.S.\ Department
of Energy under grant DE-FG03-91ER40674.

\end{document}